\documentclass[aps,prl,reprint,letterpaper,amsmath,amssymb,superscriptaddress]{revtex4-1}
\usepackage{graphicx}
\usepackage{comment}
\usepackage{color}
\usepackage{amsmath}
\newcommand{\delV}[1]{\textcolor{red}{}}
\newcommand{\addV}[1]{{#1}}

\usepackage[colorlinks,linkcolor=blue,pagecolor=blue,filecolor=blue,citecolor=blue,urlcolor=blue]{hyperref}

\begin{document}
\title{Jitter of condensation time and dynamics of spontaneous symmetry breaking in a gas of microcavity polaritons}


\author{M. V. Kochiev}
\email[]{kochievmv@mail.ru}
\affiliation{P.N. Lebedev Physical Institute, Russian Academy of
Sciences, Moscow, 119991 Russia}
\author{V. V. Belykh}
\email[]{belykh@lebedev.ru}
\affiliation{P.N. Lebedev Physical Institute, Russian Academy of
Sciences, Moscow, 119991 Russia}
\author{N. N. Sibeldin}
\affiliation{P.N. Lebedev Physical Institute, Russian Academy of
Sciences, Moscow, 119991 Russia}
\author{C. Schneider}
\affiliation{Technische Physik, Physikalisches Institut and Wilhelm Conrad R\"{o}ntgen Research Center
for Complex Material Systems, Universit\"{a}t W\"{u}rzburg, D-97074 W\"{u}rzburg, Germany}
\author{S. H\"ofling}
\affiliation{Technische Physik, Physikalisches Institut and Wilhelm Conrad R\"{o}ntgen Research Center
for Complex Material Systems, Universit\"{a}t W\"{u}rzburg, D-97074 W\"{u}rzburg, Germany}
\affiliation{SUPA, School of Physics and Astronomy, University of St Andrews, St Andrews, KY16 9SS, United Kingdom}

\date{\today}

\begin{abstract}%=======================
We investigate the statistics of microcavity polariton Bose--Einstein condensation by measuring photoluminescence dynamics from a GaAs microcavity excited by \delV{a series of} single laser excitation pulses.
We directly observe fluctuations (jitter) of the polariton condensation onset time and model them using a master equation for the \addV{occupancy} probabilities.
The jitter of the condensation onset time is an inherent property of the condensate formation and its magnitude is approximately equal to the rise time of the condensate density.
We investigate temporal correlations between the emission of condensate in opposite circular or linear polarizations by measuring \addV{the} second-order correlation function $g^{(2)}(t_1,t_2)$.
Polariton condensation is accompanied by spontaneous symmetry breaking revealed by the occurrence of random (i.e., \addV{varying} from pulse to pulse) circular and linear polarizations of the condensate emission.
The degree of circular polarization generally changes its sign in the course of condensate decay, in contrast to the degree of linear polarization.
\end{abstract}
\maketitle

\section{Introduction}%============================
\addV{A} Bose--Einstein condensate (BEC) has a macroscopic wave function that determines its quantum properties.
Since the first demonstration, BECs of microcavity (MC) polaritons~\cite{Kasprzak2006} have been investigated intensively \cite{Sanvitto2012}.
The macroscopic occupation of the ground state, the narrowing of the emission angular (momentum) distribution, temporal coherence manifesting itself in the spectral narrowing of the emission line, as well as spatial coherence, are the important criteria of condensation \cite{Deng2007, Fischer2014, Anton2014, Mylnikov2015, Demenev2016, Rozas2018}.

In the studies of the polariton BEC dynamics, the experimental data are typically averaged over a large number of excitation pulses.
However, some key phenomena are washed out upon such averaging.
This can be demonstrated by the following example.
An important property of BECs is spontaneous symmetry breaking revealed by the onset of polariton spin polarization reflected in the polarization of the MC emission.
The emission of a polariton BEC usually exhibits linear polarization pinned to one of the crystallographic axes~\cite{Klopotowski2006,Laussy2006,Shelykh2006,Kasprzak2007}, but, in the cases of very homogeneous samples, BEC emission is almost unpolarized on average.
However, it was found that, in individual events of MC emission under pulsed excitation, the \addV{degree of linear, diagonal linear, and circular polarization of luminescence from a BEC, as well as its absolute degree of polarization,} was relatively high~\cite{Baumberg2008, Read2009, Ohadi2012}, indicating spontaneous symmetry breaking. 
The dynamics of spontaneous symmetry breaking was studied with temporal resolution in Ref.~\cite{Sala2016} using a streak-camera-based photon correlation technique, introduced in Refs.~\cite{Assmann2009,Tempel2012}.

Recent studies of the MC emission statistics, allowing to reveal second- and higher-order coherence, pave the way to the further understanding of the properties and dynamics of the coherent state \cite{Assmann2009, Silva2016, KlaasPRL2018, Delteil2018, Fink2018, Sassermann2018}.
In particular, the second- and higher-order coherence of MC photoluminescence (PL) shows the crossover between thermal and coherent states~\cite{Assmann2009,Tempel2012} and is used to prove the polariton lasing regime~\cite{Kasprzak2008}.
Experiments performed using the classical Hanbury Brown and Twiss scheme made it possible to investigate the role of the lateral confinement of polaritons \cite{KlaasPRL2018}, parametric polariton scattering \cite{Sassermann2018}, and so on.
Colored cross-correlation experiments showing the antibunching of photons with different energies (i.e., of different colors) emitted by cavity polaritons \cite{Silva2016} performed with high temporal resolution generalize the Hanbury Brown--Twiss effect to the frequency domain [i.e., $g^{(2)}(\omega_1,\omega_2)$ measurements].
The spatial distribution of the BEC was shown to change from pulse to pulse~\cite{Estrecho2018}.

In the present work, we resolve the time dynamics of a polariton BEC \addV{observed after individual} excitation pulses.
Unlike most other studies, limited to the measurements of zero-delay correlations, we analyze the second-order correlation function $g^{(2)}$ of the total (in all polarizations) number of photons at two arbitrary moments of time.
These data clearly reveal fluctuations in the BEC onset time (jitter), which is studied experimentally for the first time.
This jitter affects all correlation measurements made with high temporal resolution and appropriate corrections have to be made.
By means of jitter-corrected cross-correlation measurements of the number of photons with opposite polarizations, we observe the buildup and decay of the condensate spin polarization, which is almost absent on average.
The onset of the polarization precedes that of the ground-state occupancy.
The dynamics of the degrees of circular and linear polarization are different: circular polarization decays earlier than linear and generally changes its sign.

\section{Experimental details}%========================
The sample under study is a \addV{$3/2\lambda$} MC with 12 GaAs/AlAs quantum wells of width 7~nm and \addV{top and bottom} Bragg reflectors made of 32 and 36 AlAs/Al$_{0.13}$Ga$_{0.87}$As pairs, \addV{respectively}.
It has a $Q$ factor of about 7000 and a Rabi splitting of 5~meV \cite{Belykh2013,Mylnikov2015}.
The experiments were performed at a temperature of $T = 10$~K. The results presented below correspond to a photon--exciton detuning of \addV{$\Delta = -5$}~meV; however, we performed experiments for detunings up to \addV{$1$}~meV, which showed qualitatively similar results.
The sample was mounted in a cold-finger cryostat and excited by radiation from a mode-locked Ti-sapphire laser generating a periodic ($f = 76$~MHz) train of 2.5-ps-long pulses at the wavelength corresponding to the minimum of the mirror reflectivity, 13--17~meV above the bare exciton energy.
The laser beam was focused into a 20-$\mu$m spot on the sample surface.
The PL emitted within $15^\circ$ around the sample normal was collected with a 0.25-NA microobjective
%(10x)
and split by a Wollaston prism into two beams with orthogonal linear polarizations.
In this way, the PL spot at the sample surface was transformed into two orthogonally linearly polarized spots imaged with a magnification of 3.3 onto the slit of a Hamamatsu streak camera operating with 3~ps time resolution.
For circular-polarization-resolved measurements, a $\lambda/4$ plate was installed in the optical path before the Wollaston prism.
A cylindrical lens was used to spread the luminescence spots along the slit to avoid streak camera saturation at high excitation powers.
The repetition rate of the laser pulses was lowered to 25~Hz by an acousto-optical pulse picker to match the frame rate of the streak-camera CCD, which is limited by \addV{the CCD} readout time. 
Pulse picking was synchronized with the CCD to record a single emission pulse per frame.
An example of streak camera images corresponding to a single emission event divided into left- and right-handed circular polarizations is presented in Fig.~\ref{fig:cw}(a), and a streak image accumulated over 20000 emission pulses is presented in Fig.~\ref{fig:cw}(b).

\section{Results and discussion}%============================

The emergence of spatial coherence in the course of condensate formation, as well as \addV{the persistence of} the polariton \addV{type of} dispersion above the \addV{condensation} threshold for the sample under study were demonstrated in Refs.~\cite{Belykh2013,Mylnikov2015}.
Photoluminescence spectra of the MC below and above \addV{the} condensation threshold at different \addV{detection} angles are shown in Supplemental Material \cite{Suppl}.
Here, we concentrate on the statistical properties of the condensate dynamics recorded in single MC excitation events.

\subsection{Time-integrated properties}%=========================
\label{sec:cw}

\begin{figure}[ht]
\begin{center}
\includegraphics[width=0.9\columnwidth]{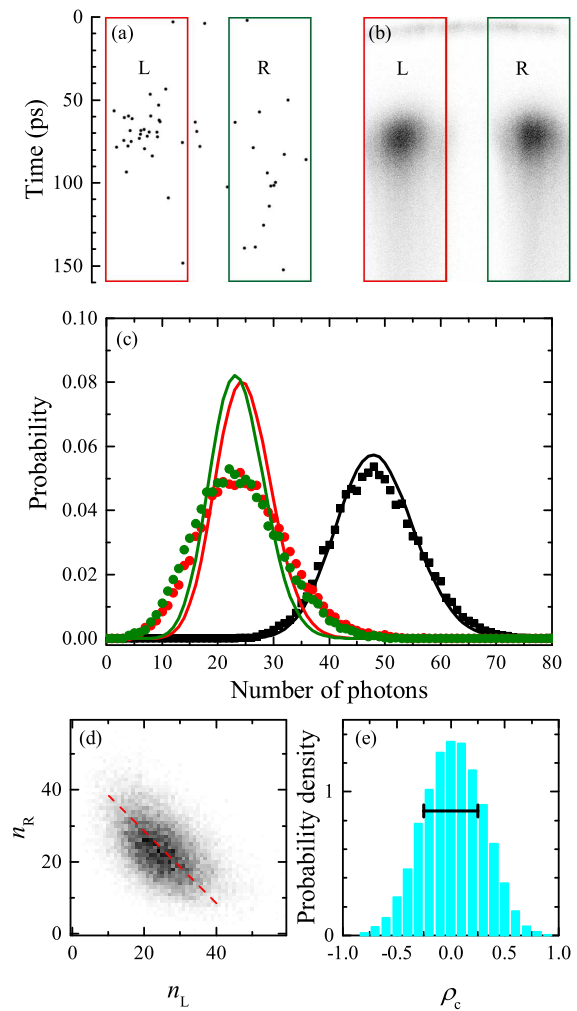}
\caption{Streak camera images obtained in a single pulse (a) and accumulated over 20000 pulses (b) for the two opposite circular polarizations at $P=2.5P_\text{thr}$.
Dots corresponding to the detected photons are enlarged in (a) for better visibility.
The horizontal axis \addV{represents the} coordinate along the \addV{streak-camera} photocathode, which corresponds to the convolution of the spatial and angular (due to the cylindrical lens) distributions of \addV{the emission intensity}.
(c) Separate distributions for left and right circular polarizations (red and green circles, respectively) and the total number of photons (squares).
Solid lines show Poisson distributions with the corresponding mean values.
(d) Scatter plot for two circular polarizations.
\addV{The red dashed line corresponds to a constant total number of photons $n=n_\text{R}+n_\text{L}$.}
(e) Probability density distribution for the degree of circular polarization $\rho_\text{c}$.
The horizontal dash shows the distribution width $\sqrt{\langle \rho_{\text em}^2 \rangle}$ corrected by excluding the Poissonian contribution (see text).} \label{fig:cw}
\end{center}
\end{figure}

First we discuss the time-integrated characteristics of the MC emission in single-pulse experiments.
We note that the MC PL averaged over a large number of excitation pulses exhibits no circular polarization and is slightly linearly polarized above the polariton BEC formation threshold $P_\text{thr}$.
Under horizontally polarized laser excitation, the degree and direction of linear polarization of the PL depends on the spot location \addV{at the sample surface} and on the pump power and is maximum just slightly above $P_\text{thr}$.
The statistical distribution of the total number $n$ of photons recorded in each PL pulse for an excitation power of $P=2.5 P_\text{thr}$ (where $P_\text{thr}=0.2$~mW at \addV{a repetition rate of} 4.75~MHz ) is shown in Fig.~\ref{fig:cw}(c) by black squares.
The spread in the number of detected photons is mostly determined by the low detection efficiency: the quantum yield of the photocathode is about 4\% at the \addV{PL} wavelength, and some \addV{fraction} of light is lost when passing optics and cryostat window.
\addV{For this reason,} the distribution of detected photons \addV{is} close to a Poissonian one $P(n,\lambda) = \lambda^n \exp(-\lambda)/n!$~\cite{Klaas2018}, where $\lambda = \langle n \rangle$ is $n$ averaged over a large number of excitation events.
It is shown by the black line in Fig.~\ref{fig:cw}(c).
Here $\langle n \rangle = 48.4$, giving a standard deviation of $\sigma = \sqrt{\langle n \rangle} \approx 7$. 
Note that, the statistics of the detected photon numbers in Ref.~\cite{Klaas2018} was mostly determined by the relatively small number of polaritons in a micropillar MC, rather than by the detection efficiency (which was high).

\addV{At the same time}, the corresponding distributions for the numbers of photons $n_\text{R}$ and $n_\text{L}$ having certain circular polarizations (right- and left-handed, respectively), shown in Fig.~\ref{fig:cw}(c) by red and green circles, respectively, are wider than the Poisson distributions with the \addV{same} mean values [red and green lines in Fig.~\ref{fig:cw}(c)].
\addV{Similar} broadening \addV{is observed} for the linear polarizations.
Such a super-Poissonian statistics, observed only above the BEC threshold, reveals increased fluctuations (in comparison with that of the total photon number $n$) from pulse to pulse in both polarizations.
The fluctuations of $n_\text{L}$ and $n_\text{R}$ at $P>P_\text{thr}$ should be in antiphase, since the total number of photons $n = n_\text{R}+n_\text{L}$ exhibits no such increased fluctuations and obeys a Poisson distribution.
These fluctuations are caused by the spontaneous polarization of the BEC emission.
This is illustrated in Fig.~\ref{fig:cw}(d), showing the \addV{scatter plot for the values of} $n_\text{R}$ and $n_\text{L}$ \addV{measured in individual pulses}. 
The dashed line in this plot corresponds to a constant total number of photons $n=n_\text{R}+n_\text{L}$.
The increased spread along the direction of this line is mostly determined by the antiphase fluctuations in $n_\text{R}$ and $n_\text{L}$.
On the other hand, the spread in the perpendicular direction is mainly \addV{caused} by the Poissonian fluctuations in $n$.

Now we \addV{aim} to determine the statistics \addV{of} the \addV{degree of} polarization.
Usually, \addV{this quantity} is defined as
\begin{equation}
\rho_\text{det} = \frac{(n_2 - n_1)} {(n_2 + n_1)},
\label{eq:detPol}
\end{equation}
where $n_1$ and $n_2$ stand for the \emph{detected} photon numbers in opposite polarizations (circular or linear)\cite{Baumberg2008,Ohadi2012}.
However, the spread in the values of $\rho_\text{det}$ defined in such a way is largely contributed by the Poisson fluctuations in $n_1$ and $n_2$ due to the low detection efficiency $\alpha \ll 1$.
The real polarization degree of the \emph{emitted} photons is defined as
\begin{equation}
\rho_\text{em} = \frac{N_2 - N_1}{N_2 + N_1},
\label{eq:Pol}
\end{equation}
where $N_1$ and $N_2$ are corresponding numbers of emitted photons (numbers of polaritons at the bottom of the lower polariton branch), so that $\langle n_1 \rangle = \alpha \langle N_1 \rangle$ and $\langle n_2 \rangle = \alpha \langle N_2 \rangle$.
We are interested in $\sqrt{\langle \rho^2_\text{em} \rangle}$.
Taking into account that $N = N_1+N_2$ is almost constant from pulse to pulse [otherwise, we would observe significant deviation from the Poisson distribution for $n$ in Fig.~\ref{fig:cw}(d)] and, according to Eq.~(\ref{eq:n12}) in the Appendix, $\langle N_{1,2}^2 \rangle = (\langle n_{1,2}^2 \rangle - \langle n_{1,2} \rangle) / \alpha^2$, we have:
\begin{equation}
\sqrt{\langle \rho_\text{em}^2 \rangle} = \sqrt{\frac{\langle (n_1-n_2)^2 \rangle}{\langle n \rangle^2} - \frac{1}{\langle n \rangle}}.
\label{eq:Pol2}
\end{equation}
Here $n = n_1 + n_2$.
In our case, the probability distribution of the detected degree of circular polarization, defined according to Eq.~\eqref{eq:detPol} and shown in Fig.~\ref{fig:cw}(e), has a spread of $\sqrt{\langle \rho_\text{det}^2 \rangle} = 0.29$. 
However, the real polarization spread of the emitted photons according to Eq.~\eqref{eq:Pol2} is smaller: $\sqrt{\langle \rho_\text{em}^2 \rangle} = 0.25$.
The probability distribution is calculated as the number of emission pulses with \addV{the} degree of circular polarization between $\rho_\text{c}-\Delta \rho_\text{c}$ and $\rho_\text{c}+\Delta \rho_\text{c}$ normalized to the total number of emission pulses and $2\Delta \rho_\text{c}$, where we select $2\Delta \rho_\text{c}=0.1$. 
\addV{We note that the absolute degree of} polarization above the threshold is \addV{noticeably lower than} unity due to polariton--polariton interactions (see Ref.~\cite{Read2009}). 

These indications of spontaneous polarization buildup are observed only in the BEC regime. In the absence of a condensate, MC emission shows completely Poissonian distributions of $n_1$ and $n_2$ (see the Supplemetal Material \cite{Suppl}).%????????????????????

\subsection{Second-order correlation function}%=====================
Only \addV{a} few photons or tens of photons per pulse can be detected in our setup [Fig.~\ref{fig:cw}(a)].
In order to obtain time resolution, each image [such as that in Fig.~\ref{fig:cw}(a)] is divided into time bins of 5~ps, in which the number of photons becomes even smaller, so \addV{that mean number} may be less than 1.
A statistical approach is needed to \addV{analyze this kind of} data, and, therefore, we determine the second-order correlation function \cite{Glauber1963}
\begin{multline}
g^{(2)}(n_1(t_1),n_2(t_2)) = \frac{\langle \hat a^\dag(t_1) \hat a^\dag(t_2) \hat a (t_2) \hat a (t_1) \rangle}{\langle \hat a^\dag(t_1) \hat a (t_1) \rangle \langle \hat a^\dag(t_2) \hat a (t_2)  \rangle}  \\ =
\frac{\langle n_1(t_1) n_2(t_2) \rangle}{\langle n_1(t_1) \rangle \langle n_2(t_2) \rangle},
\label{eq:g2}
\end{multline}
where $\hat a^\dag$ and $\hat a$ are the photon creation and annihilation operators.
It is shown in the Appendix that the Poissonian noise (which affects the distributions of $n_1$ and $n_2$ due to the detection process) has no impact on $g^{(2)}$ unless one calculates autocorrelation for $t_1 = t_2$, so that
\begin{equation}
g^{(2)}(N_1(t_1),N_2(t_2)) = g^{(2)}(n_1(t_1),n_2(t_2)),
\label{eq:gnN12}
\end{equation}
\begin{multline}
g^{(2)}(N_1(t_1),N_1(t_2)) = \\
g^{(2)}(n_1(t_1),n_1(t_2)) - \delta_{t_1,t_2} / n_1(t_1).
\label{eq:gnN}
\end{multline}
Here, $\delta$ is the Kronecker delta. It is convenient to represent the numerator of (\ref{eq:g2}) via centered variables:
\begin{multline}
g^{(2)}(N_1(t_1),N_2(t_2)) = 1 + \frac{\langle \Delta N_1 (t_1) \Delta N_2 (t_2) \rangle}
{\langle N_1(t_1) \rangle \langle N_2(t_2) \rangle},
\label{eq:g2b}
\end{multline}
where $\Delta N_{1,2}(t) = N_{1,2}(t) - \langle N_{1,2}(t) \rangle$.
Independent fluctuations of $N_1$ and $N_2$ lead to $g^{(2)} = 1$, correlated fluctuations lead to $g^{(2)} > 1$, and anticorrelated fluctuations, which take place, e.g., \addV{in the case of} spontaneous polarization buildup, lead to $0 \le g^{(2)} < 1$.
The case of $g^{(2)} = 0$ corresponds to completely polarized emission.

\subsection{Fluctuations of the BEC formation time}%=================================
\label{sec:jitter}
\begin{figure*}
\begin{center}
\includegraphics[width=11cm]{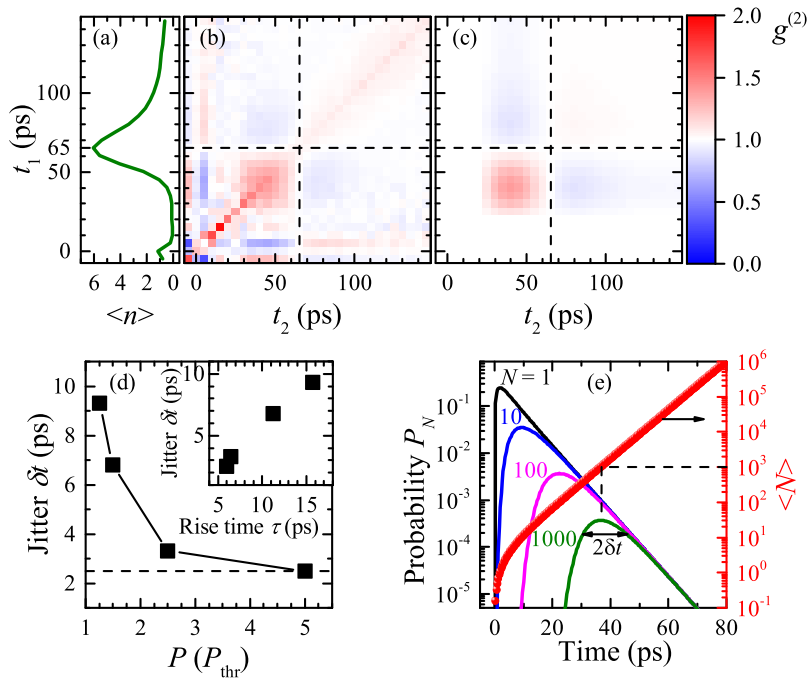}
\caption{(a) Time dependence of the total number $\langle n \rangle$ of detected photons averaged over 20000 pulses.
(b), (c) $g^{(2)}$ for the total (\addV{summed over} all polarizations) number $N$ of photons emitted at different times demonstrating the jitter effect, obtained from \addV{the} experimental data (b) and calculated using Eq.~\eqref{eq:g2vsJitter} (c).
The dashed lines correspond to the time of the PL maximum: $t_{1,2} = t_\text{max}=65$~ps.
(d)~Dependence of the BEC pulse jitter on the pump power above the BEC threshold.
The dashed line corresponds to the setup jitter.
Inset: BEC jitter versus the BEC rise time $\tau$; different points correspond to different excitation powers.
(e)~The calculated dynamics of the probabilities $P_N$ of having $N$ polaritons in the ground state (left scale) and the corresponding dynamics of $\langle N \rangle$ (right scale).
The dashed lines show that time of the probability maximum corresponds to $\langle N(t) \rangle = N$.}
\label{fig:jitter}
\end{center}
\end{figure*}

BEC formation is a stochastic process, and the time when the condensate emerges varies from pulse to pulse.
This leads to the jitter of PL pulses and the temporal broadening of the averaged PL dynamics.
This jitter can be revealed in the behavior of the second-order correlation function \addV{$g^{(2)}(N(t_1),N(t_2))$} of the total (\addV{i.e., summed over} all polarizations) number  of emitted photons \addV{$N = N_1 + N_2$}.
The PL dynamics at $P=2.5P_\text{thr}$ is shown in Fig.~\ref{fig:jitter}(a), and the corresponding
correlation function $g^{(2)}(N(t_1),N(t_2))$, calculated using Eq.~\eqref{eq:gnN}, is shown in Fig.~\ref{fig:jitter}(b).
The plot clearly shows four regions separated by the lines $t_1 = t_\text{max}$ and $t_2 = t_\text{max}$ (shown in the figure with the dashed lines), where $t_\text{max} = 65$~ps is the time corresponding to the PL maximum.
Indeed, as \addV{we show in the next paragraph}, jitter leads to anticorrelation [blue regions in Fig.~\ref{fig:jitter}(b)] or correlation [red regions in Fig.~\ref{fig:jitter}(b)] between $N(t_1)$ and $N(t_2)$ depending on whether $t_1$ and $t_2$ are on the different sides or on the same side of $t_\text{max}$, respectively.

\addV{We are now} going to determine the characteristic jitter time $\delta t = \langle (\Delta t)^2 \rangle^{1/2}$ related to the fluctuations of the condensation onset time \addV{in our experiments}.
Let us calculate the contribution to $g^{(2)}(N(t_1),N(t_2))$ arising from BEC time jitter \emph{only}.
We use Eq.~\eqref{eq:g2b} and assume that the \addV{kinetic dependence} $N(t)$ \addV{recorded after individual excitation pulses} differs \addV{from each other} only by a \addV{random} time offset $\Delta t$ \addV{that is} small compared with the characteristic time of the \addV{kinetics}.
Then, $\Delta N(t) \approx \langle N(t) \rangle' \Delta t$, where $'$ denotes differentiation on the time variable, and
\begin{equation}
g^{(2)}(N(t_1),N(t_2)) \approx 1+ \frac{\langle n(t_1) \rangle' \langle n(t_2) \rangle' (\delta t)^2}
{\langle n(t_1) \rangle \langle n(t_2) \rangle}
\label{eq:g2vsJitter}
\end{equation}
Here we take into account that $\langle N(t) \rangle' / \langle N(t) \rangle \approx \langle n(t) \rangle'/\langle n(t) \rangle$.
We calculated $g^{(2)}$ using Eq.~(\ref{eq:g2vsJitter}) with $\langle n(t) \rangle$ and $\langle n(t) \rangle'$ taken from the experiment and jitter time $\delta t = 7.6$~ps (the best fit value). 
\addV{The resulting plot} is shown in Fig.~\ref{fig:jitter}(c) and is very close to the experimental picture [Fig.~\ref{fig:jitter}(b)].

The jitter time $\delta t$ obtained from the fit is shown in Fig.~\ref{fig:jitter}(d) as a function of the excitation power.
As expected, jitter is maximal ($\sim 10$~ps) just above the BEC threshold and at $P = 5P_\text{thr}$ decreases to 2.5~ps, which corresponds to the instrumental jitter $\delta t_\text{instr}$.
The latter limits the setup time resolution and is obtained from the measured duration of the laser pulse.

\begin{figure*}
\begin{center}
\includegraphics[width=14cm]{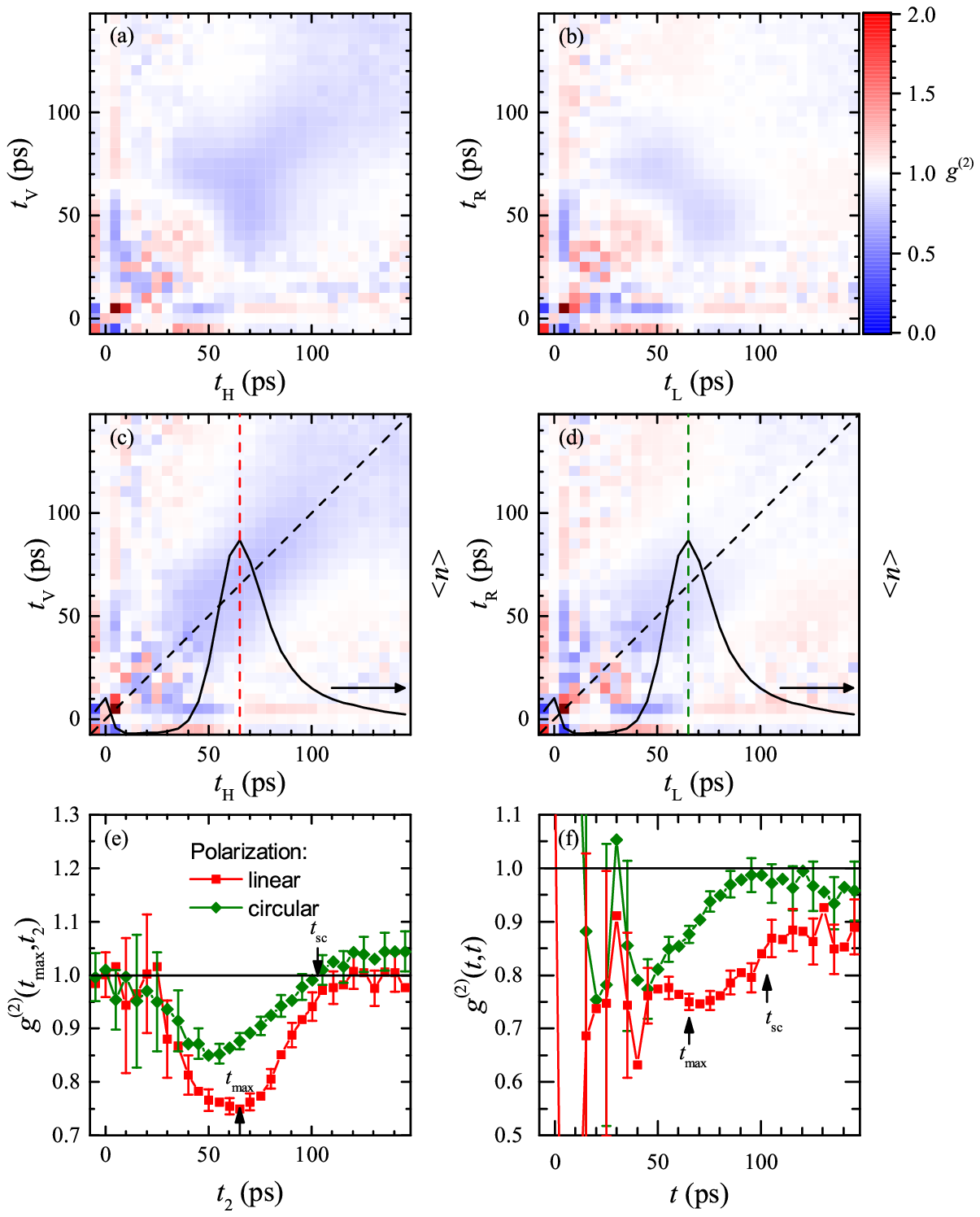}
\caption{(a),(b) Cross-correlation \addV{function} $g^{(2)}(N_1(t_1),N_2(t_2))$ \addV{for} the numbers of photons with opposite linear (a) and circular (b) polarizations.
(c),(d) The same as (a) and (b) after subtracting the jitter contribution.
The solid lines show the PL kinetics [Fig.~\ref{fig:jitter}(a)], \addV{the} dashed lines show the profiles with $t_1=t_\text{max}$ and $t_1=t_2$.
(e) Cross-correlation profiles $g^{(2)}(t_\text{max},t_2)$ for linear and circular polarizations show the relation between polarizations at time $t_\text{max}$ and at other moments $t_2$.
(f) Cross-correlation profiles $g^{(2)}(t,t)$ for linear and circular polarizations show the degree of polarization.
Arrows mark the time of the PL maximum and the time of sign reversal for the circular polarization.
Error bars in panels (e) and (f) show standard deviations.}
\label{fig:g2}
\end{center}
\end{figure*}

We can estimate theoretically the BEC formation time jitter.
The dynamics of the average polariton number $\langle N \rangle$ in the ground state is usually described using the Boltzmann kinetic equation:
\begin{equation}
\frac{d \langle N \rangle}{dt} = w (\langle N \rangle + 1) - \gamma \langle N \rangle,
\end{equation}
where $w$ is the rate of polariton scattering from the reservoir to the ground state and $\gamma$ is the total rate of polariton escape from the MC and scattering from the ground state to the reservoir.
However, here we are interested in the fluctuations of $N$ and, therefore, a description relying on a master equation, \addV{written} in terms of probabilities $P_N$ of having $N$ polaritons in a given state, is more appropriate \cite{Grundmann1997,Doan2008}:
\begin{multline}
\frac{d P_N}{dt} = w N P_{N-1} - w (N + 1) P_{N} -\\
- \gamma N P_N + \gamma (N + 1) P_{N+1}.
\label{eq:Master}
\end{multline}
Note that $\sum_{N=0}^{\infty}P_N = 1$ and $\langle N\rangle = \sum_{N=0}^{\infty} N P_N$.
The dynamics of \addV{the} calculated probabilities for $N = 1, 10, 100$, and 1000 are presented in Fig.~\ref{fig:jitter}(e) together with the dynamics of $\langle N\rangle$.
We considered only the initial condensation stage when $N$ is much lower than \addV{the} number of excitons in the reservoir and $w$ and $\gamma$ are constant (later on, $w$ decreases to \addV{the value of} $w=\gamma$, giving a maximum in the dynamics of $\langle N \rangle$, and then to $w< \gamma$, giving the decay of $\langle N \rangle$). 
Calculations are done for the values \addV{of} $w = 0.46$~ps$^{-1}$ and $\gamma = 0.3$~ps$^{-1}$.
As expected, the maximum in $P_N(t)$ corresponds to the time when $\langle N(t) \rangle = N$.
The width of \addV{the} $P_N(t)$ kinetic dependencies for $N \gg 1$ is almost independent of $N$ and gives the time uncertainty \addV{$\delta t$} of the \addV{onset of} condensation. 
According to our model, the BEC formation jitter is equal, to a constant numerical factor of the order of unity, to the characteristic rise time of $\langle N \rangle$ (for $\langle N \rangle \gg 1$):
\begin{equation}
\delta t \sim 1/(w-\gamma) = \langle N \rangle / \frac{d\langle N \rangle}{dt}.
\end{equation}
This relation is confirmed experimentally by the dependence of $\delta t$ on the PL rise time $\tau = \langle N \rangle / \frac{d\langle N \rangle}{dt}$ shown in the inset in Fig.~\ref{fig:jitter}(d).

\subsection{Correlations of photons with opposite polarizations}%=====================
\addV{Let us now} analyze the cross-correlation \addV{function} $g^{(2)}(N_1(t_1),N_2(t_2))$ for the numbers of photons with \emph{opposite} polarizations $N_1$ and $N_2$ [recall that, according to Eq.~\eqref{eq:gnN12}, this is the same as $g^{(2)}(n_1(t_1),n_2(t_2))$].
The correlation \addV{functions} for \addV{the numbers of emitted photons with the} opposite linear and circular polarizations are shown in Figs.~\ref{fig:g2}(a) and \ref{fig:g2}(b), respectively.
We have already shown that $g^{(2)}$ for the total number of photons $N=N_1+N_2$ is almost \addV{entirely} determined by the fluctuations in the BEC formation time (jitter), which are \addV{on} the order of the PL rise time.
Thus, we assume that jitter plays an important role in $g^{(2)}(N_1(t_1),N_2(t_2))$ as well.
We have to distinguish between two independent contributions to $\Delta N_{1,2}$ in Eq.~(\ref{eq:g2b}): the total jitter of $N = N_1 + N_2$ (like ``motion of the center of mass'') and spontaneous polarization (``internal motion''), independent of the former.
We can write $\Delta N_{1,2} = \Delta N_{1,2}^\text{jit} + \Delta N_{1,2}^\text{sp}$,
where $\Delta N_{1,2}^\text{jit} \approx \langle N_{1,2} \rangle ' \Delta t$ is the contribution from jitter in the total photon number (and  $\Delta t$ is the corresponding jitter time) and
$\Delta N_{1,2}^\text{sp}$
 is the contribution due to spontaneous polarization.
Thus,
\begin{multline}
g^{(2)}(N_1(t_1),N_2(t_2)) = \\ =1 + \frac{\langle N_1(t_1) \rangle' \langle N_2(t_2) \rangle' (\delta t)^2 +
\langle \Delta N_1^\text{sp} (t_1) \Delta N_2^\text{sp} (t_2) \rangle}
{\langle N_1(t_1) \rangle \langle N_2(t_2) \rangle}.
\label{eq:g2c}
\end{multline}
Taking \addV{the} jitter values determined in the previous section, we can exclude the jitter contribution from $g^{(2)}$ as is shown in Figs.~\ref{fig:g2}(c) and \ref{fig:g2}(d) for opposite linear and circular polarizations, respectively.
One can see that, after this correction, the values of $g^{(2)}$ that are less than unity are concentrated mostly along the diagonal for both circular and linear polarizations.

We will consider two types of profiles (cross sections) of such 2D plots [Figs.~\ref{fig:g2}(c) and \ref{fig:g2}(d)]: vertical (or horizontal) and diagonal.
For a vertical profile, we fix one time variable at the value corresponding to the PL maximum [vertical dashed lines in Figs.~\ref{fig:g2}(c) and \ref{fig:g2}(d)].
The values of $g^{(2)}(N_1(t_\text{max}),N_2(t_2))$ reflect the relation between the polarization at a given time and the polarization at the PL maximum.
The dependence $g^{(2)}(N_1(t_\text{max}),N_2(t_2))$ for the circular polarization\delV{and $t_\text{L} = 65$~ps (where PL reaches its maximum)} is shown in Fig~\ref{fig:g2}(e) by green diamonds.
The correlation function $g^{(2)}$ drops below unity at the \addV{beginning} of the condensation process and attains a minimum near $t_2 = t_\text{max}$.
Over almost the entire time range where the BEC exists, $g^{(2)}<1$, indicating that the spontaneous polarization of the PL remains of the same sign, i.e., in the same direction on the Poincar\'e sphere.
However, \addV{at} the PL tail ($t_2 \ge 100$~ps), $g^{(2)}$ becomes larger than unity, indicating that circular polarization \addV{typically} changes its sign.
Such a behavior was also observed for pillar MCs in Ref.~\cite{Sala2016}.
The corresponding profile for the linear polarization [red squares in Fig.~\ref{fig:g2}(e)] behaves similarly to that for the circular polarization.
However, $g^{(2)}$ \addV{only} approaches unity \addV{from below without crossing this} level \addV{at the tail of the kinetics}.
Thus, linear polarization generally \addV{exhibits no} sign \addV{reversal} over the investigated time range, but ``forgets'' its initial direction at the tail of the \addV{kinetics}. 
This means that circular polarization generally changes from left- to right-handed (or vise versa) as PL evolves from its maximum \addV{towards} the tail, while linear polarization becomes randomly directed at the kinetics tail regardless of its direction at the PL maximum.

Next, we consider the diagonal profiles, where $t_1 = t_2 = t$ [diagonal dashed lines in Figs.~\ref{fig:g2}\addV{(c)} and \ref{fig:g2}\addV{(d)}].
The value of $g^{(2)}(N_1(t),N_2(t))$ allows us to track the dynamics of the \addV{degree of} polarization: spontaneous polarization drives \addV{this correlation function} below $g^{(2)}=1$, \addV{and, as it}
 is shown in the Appendix, the deviation of $g^{(2)}$ from unity reflects the degree of spontaneous polarization [see \eqref{eq:g2v}].
The dynamics of the correlation \addV{function} $g^{(2)}(N_1(t),N_2(t))$ for opposite linear and circular polarizations is shown in Fig.~\ref{fig:g2}(f).
When condensation sets in, which is indicated by an increase in the PL intensity, $g^{(2)}$ becomes less than unity both for linear and circular polarizations, indicating the emergence of PL spontaneous polarization.
The decay of circular polarization begins even before the PL reaches its maximum, and, after $\sim 50$~ps of decay, $g^{(2)}$ for circular polarization returns to unity.
Meanwhile, for linear polarization, $g^{(2)}$ decays \addV{later}, remaining smaller than unity even when the PL intensity have already decreased significantly.
This long-lived behavior of $g^{(2)}$, and, thus, of the degree of linear polarization, is similar to that of spatial coherence and polariton momentum distribution reported in Refs.~\cite{Belykh2013,Mylnikov2015,Demenev2016,Demenev2018}.

As expected, $g^{(2)}(N_1(t),N_2(t))$ shows no systematic deviation from unity in the absence of condensation (see Supplemental Material \cite{Suppl}).

\section{Conclusions}%============================
We have studied fluctuations in the onset time \addV{of Bose condensation} and \addV{in the degree of linear and circular} spontaneous polarization in a MC polariton system by measuring the temporal dynamics of the second-order correlation function $g^{(2)}$.
\addV{The analysis of} correlations between the number of photons detected at different moments in time \addV{has given} clear \addV{evidence of} jitter in the BEC onset time.
This jitter \addV{is} an inherent property of the BEC formation process \addV{resulting from} its stochastic nature.
\addV{Both the experimental data} and a simple master equation model \addV{indicate} that the \addV{magnitude of} jitter is proportional to the characteristic rise time of the \addV{polariton} ground state occupancy.
This jitter should be taken into account when interpreting the results of correlation measurements.

\addV{Measurements of the cross-correlation function} $g^{(2)}(n_1(t),n_2(t))$ \addV{for} the intensities of PL with opposite polarizations \addV{have made it possible to investigate} the dynamics of the \addV{degree of} polarization \addV{(or, more strictly, its pulse-to-pulse variance)}.
Spontaneous linear polarization \addV{of the condensate} lasts for at least 100~ps, whereas spontaneous circular polarization decays earlier and \addV{its sign is typically reversed} at the tail of the PL kinetics.

\section{acknowledgments}%$$$$$$$$$$$$$$$$$$$$$$$$$$$$$$$$$$$$
\begin{acknowledgments}
We are grateful to V.~D. Kulakovskii for his attention, support, and precious recommendations; to M.~L. Skorikov for careful reading of the manuscript and valuable remarks and advices; and to N.~A. Gippius, S.~G. Tikhodeev, and V.~A. Tsvetkov for useful discussions.
This work was supported by the Russian Foundation for Basic Research (project No. 18-02-01143) and the State of Bavaria.
\end{acknowledgments}
%=============
\appendix*
\section{APPENDIX}

\setcounter{equation}{0}
Here, we express the correlation \addV{function for} the numbers of photons \emph{emitted} by the microcavity (MC) \addV{in terms of} the numbers of \emph{detected} photons.
Let $N_1$ and $N_2$ be the numbers of photons {emitted} by the MC \delV{between which we are going to calculate the correlation,} and
$n_1$ and $n_2$ be the corresponding numbers of {detected} photons, so that $\langle n_1 \rangle = \alpha \langle N_1 \rangle$, $\langle n_2 \rangle = \alpha \langle N_2 \rangle$, where $\alpha \ll 1$ is the detection probability and averaging is done over \addV{a series of} excitation events. Indices 1 and 2 may correspond to different polarizations and/or different time moments.

We are looking for the correlation \addV{function}
\begin{equation}
g^{(2)}(N_1,N_2) = \frac{\langle N_1 N_2 \rangle}{\langle N_1 \rangle \langle N_2 \rangle}
\label{eq:RA}
\end{equation}
and \addV{are} going to express it through $n_1$ and $n_2$.

Let $F(N_1, N_2)$ be the distribution function of $N_1$ and $N_2$, with $\sum_{N_1, N_2}F( N_1, N_2) = 1$. The distribution $F(N_1, N_2)$ determines the correlation properties of $N_1$ and $N_2$, including spontaneous polarization. Then,
\begin{subequations}
\begin{align}
\langle  N_1  N_2\rangle = \sum_{ N_1, N_2}F(N_1,N_2)N_1 N_2,\\
\langle N_1^2\rangle = \sum_{N_1,N_2}F(N_1,N_2)N_1^2.
\end{align}
\end{subequations}

For a given number $N_1$ of emitted photons, the probability distribution for the number of detected photons $n_1$ is a Poissonian $P(n_1,\lambda)$ with a mean value of $\lambda = \alpha N_1$. Then, for the averages $\langle n_1 n_2\rangle$ and $\langle n_1^2 \rangle$ we have:
\begin{multline}
\langle n_1 n_2\rangle =\\ \sum_{N_1,N_2,n_1,n_2}F(N_1,N_2)P(n_1, \alpha N_1)P(n_2, \alpha N_2)n_1 n_2 = \\
=\alpha^2\sum_{N_1,N_2}F(N_1,N_2)N_1 N_2 = \alpha^2 \langle N_1 N_2 \rangle
\label{eq:n1n2}
\end{multline}
and
\begin{multline}
\langle n_1^2\rangle = \sum_{N_1,N_2,n_1}F(N_1,N_2)P(n_1, \alpha N_1)n_1^2 = \\
=\sum_{N_1,N_2}F(N_1,N_2)(\alpha^2 N_1^2 + \alpha N_1) = \\
=\alpha^2 \langle N_1^2 \rangle + \alpha \langle N_1 \rangle,
\label{eq:n12}
\end{multline}
where \addV{we took into account that} for the Poisson distribution $P(k,\lambda)$, $\langle k^2 \rangle = \lambda^2 + \lambda$.

Substituting $\langle N_1 N_2 \rangle$ and $\langle N_1^2 \rangle$ from Eqs.~\eqref{eq:n1n2} and \eqref{eq:n12} into Eq.~\eqref{eq:RA}, we obtain the final formulas for the correlation functions we are looking for:
\begin{subequations}
\label{eq:G2g2}
\begin{eqnarray}
g^{(2)}(N_1,N_2) = g^{(2)}(n_1,n_2),\label{eq:G2g2a}\\
g^{(2)}(N_1,N_1) = g^{(2)}(n_1,n_1) - 1/ \langle n_1 \rangle.\label{eq:G2g2b}
\end{eqnarray}
\end{subequations}
\delV{We will use Eq.~(\ref{eq:G2g2a}) in all cases but $N_1 \equiv N_2$ (i.e. for the zero-delay autocorrelation), when Eq.~(\ref{eq:G2g2b}) is needed.}

Next, we are going to \addV{find} how the zero-delay cross-correlation \addV{function} $g^{(2)}(N_1(t),N_2(t))$ relates to the \addV{degree of} spontaneous polarization.
According to the experimental data, \addV{there are no significant fluctuations in} the total number of emitted photons \addV{per pulse} (see Fig.~\ref{fig:cw}(c)), while $N(t)$ (for a given \addV{time} $t$) fluctuates due to the jitter effect.
Instead of $\rho_\text{em}$, defined by Eq.~\eqref{eq:Pol}, let us describe \addV{the degree of} polarization and its variance in terms of a modified variable
\begin{equation}
\tilde \rho_\text{em}(t) = {\frac{ N_1(t)-N_2(t) }{ \langle N(t) \rangle }} .
\end{equation}
Evidently, $\tilde \rho_\text{em}(t) = \rho_\text{em}(t) $ in the absence of jitter. Then, the variance is
\begin{multline}
\langle \tilde \rho_\text{em}^2(t) \rangle={\frac{ \langle (N_1(t)-N_2(t))^2 \rangle }{ \langle N(t) \rangle^2 }} = \\
= \langle \tilde \rho_\text{det}^2(t) \rangle - \frac{1}{\langle n(t) \rangle},
\end{multline}
similarly to Eq.~\eqref{eq:Pol2}.
Here $ \tilde \rho_\text{det}(t) = (n_1(t)-n_2(t))/ \langle n(t) \rangle$.
It is easy to show, that
\begin{multline}
 \langle \tilde \rho_\text{em}^2(t) \rangle = { \langle \tilde \rho_\text{em}(t) \rangle^2 +  4\langle \Delta_\text{sp}^2(t)\rangle / \langle N(t) \rangle^2} ,
\end{multline}
where $\Delta N_{1}^\text{sp}(t) = \Delta_\text{sp}(t)$ and $\Delta N_{2}^\text{sp}(t) = - \Delta_\text{sp}(t)$.
Then, we can write the polarization part of Eq.~\eqref{eq:g2c} as
\begin{multline}
g^{(2)}(N_1(t),N_2(t)) = 1 - \frac{ \langle \Delta_\text{sp}^2 (t) \rangle}
{\langle N_1(t) \rangle \langle N_2(t) \rangle} =\\
=1- \frac{\langle \tilde \rho_\text{em}^2(t) \rangle - \langle \tilde \rho_\text{em}(t) \rangle ^2}  {1-\langle \tilde \rho_\text{em}(t) \rangle ^2}=\\
=1- \frac{\langle \Delta \tilde \rho_\text{em}^2(t) \rangle}{1-\langle \tilde \rho_\text{em}(t) \rangle ^2},
\end{multline}
where $\Delta \tilde \rho_\text{em}(t) = \tilde \rho_\text{em}(t) - \langle \tilde \rho_\text{em}(t) \rangle$.
For zero average \addV{degree of} polarization \addV{$\langle \rho_\text{em}(t) \rangle = 0$} (which is the case, e.g., for circular polarization)
\begin{equation}
\langle \tilde \rho_\text{em}^2(t) \rangle = 1 - g^{(2)}(N_1(t),N_2(t))
\label{eq:g2v}
\end{equation}
So, the deviation of $g^{(2)}$ from unity reflects the degree of spontaneous polarization.

\clearpage

\section{SUPPLEMENTAL MATERIAL}

\setcounter{figure}{0}

\section{Photoluminescence spectra}

\renewcommand{\thefigure}{S\arabic{figure}}

\begin{figure*}
\begin{center}
\includegraphics[width=12cm]{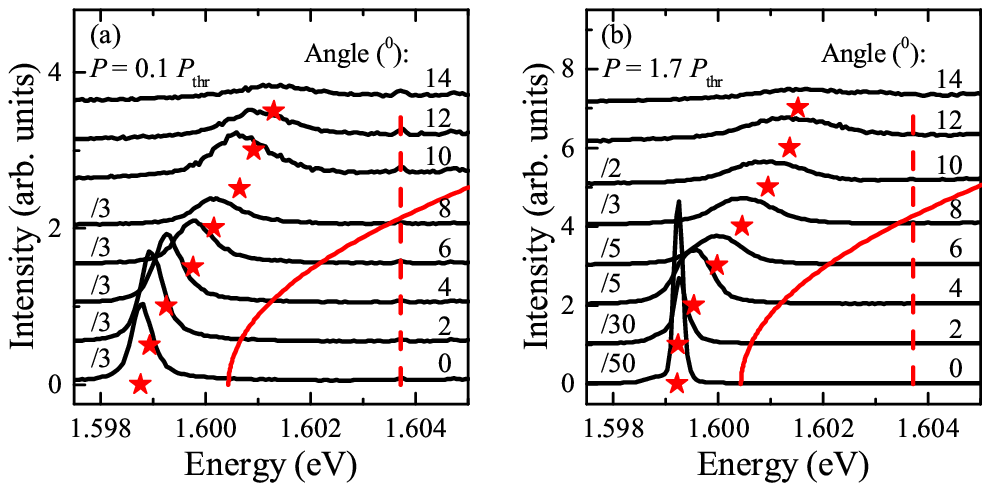}
\caption{Photoluminescence spectra of the MC at pump power below (a) and above (b) the condensation threshold recorded at different detection angles.
Stars mark the positions of the lower polariton branch PL maxima.
Red solid and dashed lines denote the cavity mode and the bare exciton energy, respectively.}
\label{fig:suppl1}
\end{center}
\end{figure*}

Photoluminescence (PL) time-integrated spectra of the MC at different detection angles, corresponding to  different polariton wavevectors, are shown in Fig.~\ref{fig:suppl1}.
We use pulsed excitation with pump power $P=0.1P_\text{thr}$ [Fig.~\ref{fig:suppl1}(a)] and $P=1.7P_\text{thr}$ [Fig.~\ref{fig:suppl1}(b)].
The detuning between the exciton and photon modes is $\Delta= -1.5$~meV.

\section{Statistics in the absence of a condensate}

\begin{figure*}
\begin{center}
\includegraphics[width=\textwidth]{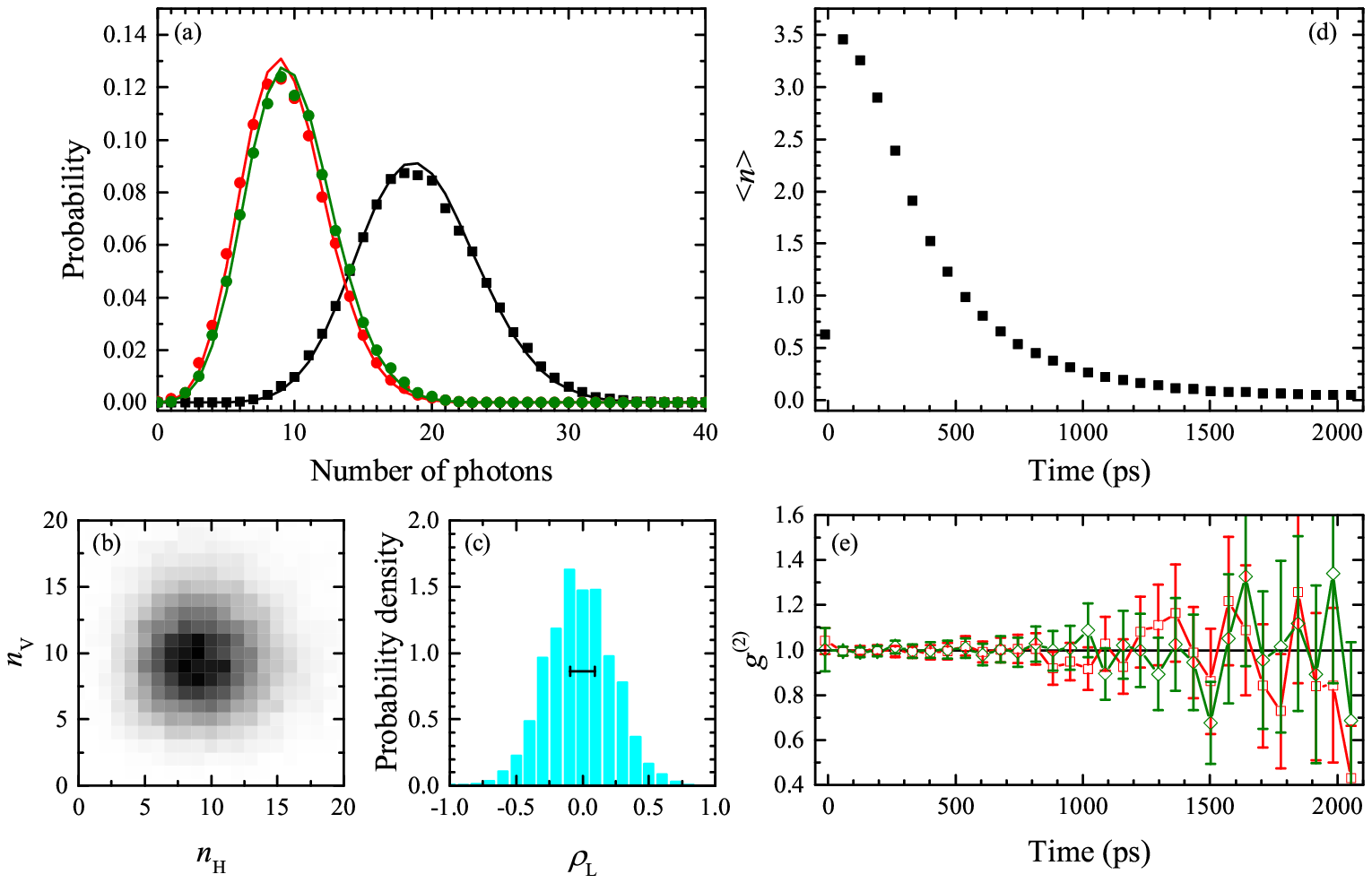}
\caption{(a) Separate distributions for the numbers of detected PL photons with the horizontal and vertical linear polarizations (red and green circles, respectively) and the total number of photons (black squares) in the absence of condensation. Solid lines show the Poisson distributions with the corresponding mean values.
(b) Scatter plot for the numbers of photons with the horizontal and vertical linear polarizations.
(c)~Probability density distribution for the degree of linear polarization $\rho_{\text L}$. 
The horizontal dash shows the distribution width $\sqrt{\langle \rho_{\text L}^2 \rangle}$ corrected by excluding the Poissonian contribution using Eq.~(3) from the main text.
(d) Time evolution of the average photon number $\langle n \rangle$.
(e) Cross-correlation function $g^{(2)}$ for the  numbers of photons with opposite linear (red squares) and circular (green diamonds) polarizations.
Error bars show standard deviations.}
\label{fig:suppl2}
\end{center}
\end{figure*}

To prove that our measurements yield values of $g^{(2)}(n_1(t_1),n_2(t_2))<1$ due to the onset of the spontaneous polarization of the BEC rather than, e.g., due to a better signal to noise ratio resulting from a dramatically increased intensity, we have also measured correlation at a very large positive detuning where the bare exciton energy corresponds to the first reflection minimum of the cavity mirrors. 
Thus, PL from the first reflection minimum is measured.
In this configuration, no BEC is formed even at high pump powers where the PL intensity is comparable to that in the BEC regime described in the main text. 
Figure~\ref{fig:suppl2}(a) shows that both the total number of photons $n$ and also the numbers $n_1$ and $n_2$ of photons detected in orthogonal linear polarizations obey Poisson distributions (and the results are the same for circular polarizations). 
Furthermore, the scatter plot of $(n_1,n_2)$ is symmetric [Fig.~\ref{fig:suppl2}(b)] as expected for uncorrelated variables. 
The distribution of the degree of linear polarization $\rho_\text{L}$ is rather broad [Fig.~\ref{fig:suppl2}(c)], but the main contribution is caused by the Poissonian noise. 
These observations are in stark contrast with those for the polariton BEC regime [Fig.~1(c),(d),(e) in the main text]. 
The PL dynamics is shown in Fig.~\ref{fig:suppl2}(d). 
It is typical of quantum wells and is much slower than that in the BEC regime. 
The second-order correlation function $g^{(2)}(n_1(t),n_2(t))$ shows no systematic deviations from unity, indicating neither linear nor circular spontaneous polarization [Fig.~\ref{fig:suppl2}(e)].

\end{document}